\documentclass[12pt]{article}

\usepackage{amsmath,amsfonts,amssymb}

\begin{document}

\begin{titlepage}

\begin{center}

\vskip 0.4 cm

\begin{center}
{\Large  \bf Hamiltonian Analysis of Mixed Derivative
Ho\v{r}ava-Lifshitz Gravity}
\end{center}

\vskip 1cm

\vspace{1em}  Josef Kluso\v{n}\footnote{Email address:
 klu@physics.muni.cz}\\
\vspace{1em}\textit{Department of Theoretical Physics and
Astrophysics, Faculty of Science,\\
Masaryk University, Kotl\'a\v{r}sk\'a 2, 611 37, Brno, Czech Republic}

\vskip 0.8cm

\end{center}

\begin{abstract}
This short note is devoted to the canonical analysis of the
Ho\v{r}ava-Lifshitz gravity with mixed derivative terms that was
proposed  in arXiv:1604.04215. We determine the algebra of
constraints and we show that there is one additional scalar degree
of freedom with respect to the non-projectable Ho\v{r}ava-Lifshitz
gravity.

\end{abstract}

\end{titlepage}
\bigskip

\newpage

\def\mP{\mathcal{P}}
\def\bn{\mathbf{n}}
\newcommand{\bC}{\mathbf{C}}
\newcommand{\bD}{\mathbf{D}}
\def\tpi{\tilde{\pi}}
\def\hf{\hat{f}}
\def\tK{\tilde{K}}
\def\bmC{\bar{\mC}}
\def\tmG{\tilde{\mG}}
\def\tPi{\tilde{\Pi}}
\def\tmC{\tilde{\mC}}
\def\tPhi{\tilde{\Phi}}
\def\tv{\tilde{v}}
\def\mC{\mathcal{C}}
\def\bk{\mathbf{k}}
\def\tp{\tilde{p}}
\def\tr{\mathrm{tr}\, }
\def\tmH{\tilde{\mH}}
\def\tPsi{\tilde{\Psi}}
\def\tY{\mathcal{Y}}
\def\nn{\nonumber \\}
\def\bI{\mathbf{I}}
\def\tmV{\tilde{\mV}}
\def\e{\mathrm{e}}
\def\bE{\mathbf{E}}
\def\bX{\mathbf{X}}
\def\bY{\mathbf{Y}}
\def\bR{\bar{R}}
\def\hN{\hat{N}}
\def\tR{\tilde{R}}
\def\hK{\hat{K}}
\def\hnabla{\hat{\nabla}}
\def\hc{\hat{c}}
\def\mH{\mathcal{H}}
\def \Gi{\left(G^{-1}\right)}
\def\hZ{\hat{Z}}
\def\bz{\mathbf{z}}
\def\bK{\mathbf{K}}
\def\iD{\left(D^{-1}\right)}
\def\tmJ{\tilde{\mathcal{J}}}
\def\tr{\mathrm{Tr}}
\def\mJ{\mathcal{J}}
\def\tk{\tilde{k}}
\def\tGamma{\tilde{\Gamma}}
\def\partt{\partial_t}
\def\parts{\partial_\sigma}
\def\bG{\mathbf{G}}
\def\str{\mathrm{Str}}
\def\Pf{\mathrm{Pf}}
\def\bM{\mathbf{M}}
\def\tA{\tilde{A}}
\newcommand{\mW}{\mathcal{W}}
\def\bx{\mathbf{x}}
\def\by{\mathbf{y}}
\def \mD{\mathcal{D}}
\newcommand{\tZ}{\tilde{Z}}
\newcommand{\tW}{\tilde{W}}
\newcommand{\tmD}{\tilde{\mathcal{D}}}
\newcommand{\tN}{\tilde{N}}
\newcommand{\hC}{\hat{C}}
\newcommand{\hg}{g}
\newcommand{\hX}{\hat{X}}
\newcommand{\bQ}{\mathbf{Q}}
\newcommand{\hd}{\hat{d}}
\newcommand{\tX}{\tilde{X}}
\newcommand{\calg}{\mathcal{G}}
\newcommand{\calgi}{\left(\calg^{-1}\right)}
\newcommand{\hsigma}{\hat{\sigma}}
\newcommand{\hx}{\hat{x}}
\newcommand{\tchi}{\tilde{\chi}}
\newcommand{\mA}{\mathcal{A}}
\newcommand{\ha}{\hat{a}}
\newcommand{\tB}{\tilde{B}}
\newcommand{\hrho}{\hat{\rho}}
\newcommand{\hh}{\hat{h}}
\newcommand{\homega}{\hat{\omega}}
\newcommand{\mK}{\mathcal{K}}
\newcommand{\hmK}{\hat{\mK}}
\newcommand{\hA}{\hat{A}}
\newcommand{\mF}{\mathcal{F}}
\newcommand{\hmF}{\hat{\mF}}
\newcommand{\hQ}{\hat{Q}}
\newcommand{\mU}{\mathcal{U}}
\newcommand{\hPhi}{\hat{\Phi}}
\newcommand{\hPi}{\hat{\Pi}}
\newcommand{\hD}{\hat{D}}
\newcommand{\hb}{\hat{b}}
\def\I{\mathbf{I}}
\def\tW{\tilde{W}}
\newcommand{\tD}{\tilde{D}}
\newcommand{\mG}{\mathcal{G}}
\def\IT{\I_{\Phi,\Phi',T}}
\def \cit{\IT^{\dag}}
\newcommand{\hk}{\hat{k}}
\def \cdt{\overline{\tilde{D}T}}
\def \dt{\tilde{D}T}
\def\bra #1{\left<#1\right|}
\def\ket #1{\left|#1\right>}
\def\mV{\mathcal{V}}
\def\Xn #1{X^{(#1)}}
\newcommand{\Xni}[2] {X^{(#1)#2}}
\newcommand{\bAn}[1] {\mathbf{A}^{(#1)}}
\def \bAi{\left(\mathbf{A}^{-1}\right)}
\newcommand{\bAni}[1]
{\left(\mathbf{A}_{(#1)}^{-1}\right)}
\def \bA{\mathbf{A}}
\newcommand{\bT}{\mathbf{T}}
\def\bmR{\bar{\mR}}
\newcommand{\mL}{\mathcal{L}}
\newcommand{\mbQ}{\mathbf{Q}}
\def\mat{\tilde{\mathbf{a}}}
\def\mtF{\tilde{\mathcal{F}}}
\def \tZ{\tilde{Z}}
\def\mtC{\tilde{C}}
\def \tY{\tilde{Y}}
\def\pb #1{\left\{#1\right\}}
\newcommand{\E}[3]{E_{(#1)#2}^{ \quad #3}}
\newcommand{\p}[1]{p_{(#1)}}
\newcommand{\hEn}[3]{\hat{E}_{(#1)#2}^{ \quad #3}}
\def\mbPhi{\mathbf{\Phi}}
\def\tg{\tilde{g}}
\newcommand{\phys}{\mathrm{phys}}

\section{Introduction}
General Relativity (GR) is one of the most beautiful physical
theories that is in perfect agreement with the current experimental
tests. On the other hand it is well known that this theory is in
conflict with the quantum mechanics since it is not perturbatively
renormalizable and hence it breaks down at high energies.  In order
to solve this problem P. Ho\v{r}ava proposed very original
formulation of theory of gravity  \cite{Horava:2009uw} which is now
known as Ho\v{r}ava-Lifshitz (HL) gravity. This theory has an
improved behavior at high energies due to the presence of the higher
order spatial derivatives in the action which implies that the
theory is not invariant under full diffeomorphism but it is
invariant under so called foliation preserving diffeomorphism
($\mathrm{Diff}_\mF$)
\begin{equation}\label{DiffF}
t'=f(t) \ , \quad x'^i=x^i(\bx,t) \ .
\end{equation}
This property offers the possibility that the space and time
coordinates have different scaling at high energies
\begin{equation}
t'=k^{-z}t \ , x'^i=k^{-1}x^i \ ,
\end{equation}
where $k$ is a constant. Consequence of this fact is that in $3+1$
dimensions the theory contains terms with $2$ time derivatives and
at least $2z$ spatial derivatives since the minimal amount of the
scaling anisotropy that is needed for the power-counting
renormalizability of this theory is $z=3$.  Then collecting all
terms that are invariant under $\mathrm{Diff}_\mF$ symmetry leads to
the general action \cite{Blas:2009qj,Blas:2010hb}
\begin{eqnarray}
S=\frac{M_p^2}{2} \int dt d^3\bx N \sqrt{g}K_{ij}
\mG^{ijkl}K_{kl}-S_V  \ ,
\end{eqnarray}
where
\begin{equation}
K_{ij}=\frac{1}{2N}(\partial_t g_{ij}-D_iN_j-D_jN_i) \ ,
\end{equation}
and where we introduced generalized De Witt metric $\mG^{ijkl}$
defined as \cite{Horava:2008ih}
\begin{equation}
\mG^{ijkl}=\frac{1}{2}(g^{ik}g^{jl}+g^{il}g^{jk})-\lambda
g^{ij}g^{kl} \ ,
\end{equation}
where $\lambda$ is an arbitrary real constant. Finally note that
$D_i$ is  the covariant derivative defined with the help of the
metric $g_{ij}$.  The action $S_V$ is the potential term action in
the form
\begin{equation}
S_V=\frac{M_p^2}{2}\int dt d^3\bx N\sqrt{g}\mV=\frac{M_p^2}{2}\int
dt d^3\bx N\sqrt{g}\left(\mL_1+\frac{1}{M^2_*}\mL_2+
\frac{1}{M^4_*}\mL_3\right) \ ,
\end{equation}
where $\mL_n$ contain all terms that are invariant under foliation
preserving diffeomorphism and where $\mL_n$ contain $2n$ derivatives
of the ADM variables $(N,g_{ij})$. In the UV when $k\gg M_*$ the
dominant contributions come from the higher derivative terms that
lead to the modified dispersion relation $m^2\propto k^6$  that
implies that this theory is power counting renormalizable. In the
opposite regime $k\ll M_*$ the dispersion relation is relativistic
and it can be shown that the theory have regions in the parameter
space where
 it is in agreement with observation.

Despite these attractive properties there is a serious problem
considering the Lorentz violation operators in the matter sector.
In fact, while the direct bounds on Lorentz violations in the gravity sector
are weak, the bounds on the Lorentz violating operators in the matter sector
are very stringent.  For that reason it is very important to prevent
Lorentz violations leaking from the gravity sector to the matter sector.

One possibility how to resolve this problem was suggested in
\cite{Pospelov:2010mp}, where the Lorentz violating gravity sector
couples to the Standard model through power suppressed operators.
However it turns out that this generic mechanism is not entirely
successful in case of HL gravity due to the fact that non-dynamical
vector gravitons are not modified with respect to GR which leads to
the quadratic divergences that should be fine-tuned away. It was
proposed in \cite{Pospelov:2010mp} to include a single term $D_i
K_{jk}D^iK^{jk}$ to the action. The presence of this term modifies
the vector graviton sector at linear order while leaves the tensor
and scalar dispersion relation qualitatively unchanged. This
proposal was further studied in \cite{Colombo:2014lta} where the
contributions of all terms of the form $(D_i K_{jk})^2$ were
analyzed. It was shown there that  all dispersion relations in the
UV now become of the type $\omega^2\propto k^4$ that are not enough
for the standard power-counting renormalizability of HL gravity.
However then it was argued in \cite{Colombo:2014lta}
 that in the presence of the mixed derivative
terms the power-counting relation is modified and these dispersion
relations provide sufficient momentum suppressions in the
amplitudes. These mixed derivative extensions were further analyzed
in \cite{Colombo:2015yha} where the new class of Lifshitz-like
scalar theories that are power-counting renormalizable and unitary
were introduced. In \cite{Coates:2016zvg} this construction was
extended to the case of HL gravity where the most general mixed
derivative form of HL was proposed. However a careful perturbative
analysis performed there shows that the presence of the new mixed
derivative terms has a dramatic impact on the theory since they
generate new degree of freedom. This is very interesting fact that
certainly deserves to be analyzed further. Natural mechanism how to
identify the number of degrees of freedom is  to perform Hamiltonian
analysis of this theory and  this is exactly the goal of this paper.
It turns out that this analysis is rather straightforward when we
introduce appropriate auxiliary fields in order to replace mixed
derivative terms with ordinary ones. The presence of these auxiliary
fields then imply new second class constraints that can be solved
for them at least in principle. In other words  terms like $(D_i
K_{jk})^2$ do not generate new dynamical degree of freedom. In fact,
we show that this scalar degree of freedom is related to the
existence of the time derivative of the lapse $N$.

The structure of this note is as follows. In the next section
(\ref{second}) we introduce mixed derivative HL gravity following
\cite{Coates:2016zvg}. Then in section (\ref{third}) we find
Hamiltonian formulation of this theory and determine number of
degrees of freedom. In Summary (\ref{fourth}) we outline our
results.
\section{Mixed Derivative HL Gravity }\label{second}
In order to construct mixed derivative HL gravity new additional
part  to the action  was introduced in \cite{Coates:2016zvg}
\begin{equation}
S_\kappa=\frac{M_p^2}{2M^2_*}\int dt d^3\bx N\sqrt{g}\mL_\kappa \
\end{equation}
that contains all  $\mathcal{D}iff_F$ invariant operators that
involve two spatial and two time derivatives. Generally the number
of independent operators is of order $10^2$. However we focus on
terms that contribute to the quadratic action. Then we consider
following contend of HL gravity
\begin{eqnarray}
\mL_1&=&2\alpha a_i a^i+\beta R \ , \nonumber \\
\mL_2&=&\alpha_1 R D_i a^i+\alpha_2 D_i a_j D^i a^j+\beta R_{ij}R^{ij}+
\beta_2 R^2 \ , \nonumber \\
\mL_3&=&\alpha_3 D_i D^i RD_j a^j+\alpha_4 D^k D_k a_i D_j D^j a^i\nonumber \\
&+&\beta_3 D_i R_{jk} D^i R^{jk}+\beta_4 D_i R D^i R \nonumber \\
\end{eqnarray}
while the mixed derivative terms have the form
\begin{eqnarray}
\mL_\kappa&=&D_i K_{jk}D_l K_{mn}M^{ijklmn}\nonumber \\
&+&2(\sigma_1 \mA_i \mA^i+\sigma_2 \mA_i D^i K+\sigma_3 \mA_i D_j K^{ij}) \ ,
\nonumber \\
\end{eqnarray}
where
\begin{eqnarray}
M^{ijklmn}&=&\gamma_1 g^{ij}g^{lm}g^{kn}+\gamma_2 g^{il}g^{jm}g^{kn}+
\gamma_3 g^{il}g^{jk}g^{mn}+\gamma_4 g^{ij}g^{kl}g^{mn} \ ,  \nonumber \\
\end{eqnarray}
and where
\begin{equation}
\mA_i=\frac{1}{2N}(\partial_t a_i-N^j D_ja_i-a_j D_i N^j)
\end{equation}
is $\mathrm{Diff}_{\mathcal{F}}$ covariant combination that contains the time
derivative of $N$. There is also
$\mathrm{Diff}_{\mathcal{F}}$ covariant combination that contains the time
derivative of the $3-$curvature
\begin{equation}
r_{ij}=\frac{1}{2N}(\dot{R}_{ij}-N^k D_k R_{ij}-R_{ik}D_j N^k-R_{jk}D_i N^k)
\end{equation}
so that terms like $K^{ij}r_{ij}$ and $Kr$ are $\mathrm{Diff}_{\mathcal{F}}$
scalars with the right number of derivatives.

Our goal is to perform the Hamiltonian analysis of this theory in
order to confirm that it contains new scalar degree of freedom.
Before we proceed to this analysis we will argue that we can
consistently ignore  terms containing $r_{ij}K^{ij}$ and $rK$. To do
this we use the formula for the variation of the Ricci tensor
\begin{equation}
\delta R_{ij}=\frac{1}{2}g^{kl}[D_kD_j \delta g_{il}
+D_k D_i \delta g_{jl}-D_i D_j
\delta g_{kl}-D_k D_l \delta g_{ij}]
\end{equation}
and hence
\begin{eqnarray}
\dot{R}_{ij}
=\frac{1}{2}g^{kl}(D_k D_j \dot{g}_{il}+D_k D_i \dot{g}_{jl}-
D_i D_j \dot{g}_{kl}-D_k D_l \dot{g}_{ij})  \ . \nonumber \\
\end{eqnarray}
Now we use the fact that
\begin{equation}
\dot{g}_{ij}=2N K_{ij}+D_i N_j+D_j N_i
\end{equation}
so that  $r_{ij}$ has the form
\begin{eqnarray}
r_{ij}&=&\frac{1}{2N}g^{kl} \left(D_kD_j (2NK_{il}+D_i N_l+D_l
N_i)+D_kD_i(2NK_{jl}+D_j N_l+D_l N_j) \right.
\nonumber \\
&-&D_iD_j (2NK_{kl}+D_kN_l+D_lN_k)-D_kD_l (2NK_{ij}+D_iN_j+D_jN_i)
\nonumber \\
&-& \left. N^k D_k R_{ij}-R_{ik}D_j N^k-R_{jk}D_i N^k\right) \ . \nonumber \\
\end{eqnarray}
Then certainly terms like $K^{ij}r_{ij}$ and $Kr$ add additional
terms in the action that contain $a_i$ and its covariant derivative
together with  covariant derivative of $K_{ij}$. As we will argue in
the next section it is natural to replace $K_{ij}$ with auxiliary
field whenever covariant derivative acts on $K_{ij}$. Then we see
that these terms contribute as  additional potential terms for these
auxiliary fields and vector $a_i$ and hence do not have an impact on
the canonical structure of the theory. Then in order to simplify
resulting analysis we will not include terms like $K^{ij}r_{ij}$ and
$Kr$ into the action.
\section{Canonical Analysis}\label{third}
The specific property of the mixed derivative form of HL gravity is
that it contains the covariant derivative of $K_{ij}$ so that when
we proceed to the Hamiltonian formalism we would get relation
between conjugate momenta $\pi^{ij}$ and differential operator of
the second order in the spatial derivatives acting on $K_{ij}$. Then
in order to invert this relation we should introduce some kernel of
this differential operator and we would find that $K_{ij}$ is given
as an integral over the second argument of this kernel multiplied by
some functions of $\pi^{ij}$. However then we would find that the
Hamiltonian is non-local functional of conjugate variables and hence
it is very difficult to show that this Hamiltonian has correct
Poisson brackets with the generators of the spatial diffeomorphism.
For that reason it is more natural to introduce auxiliary fields
$A_{ij}$ instead $K_{ij}$ in the expressions that contain covariant
derivative of $K_{ij}$.  Explicitly, we consider the action in the
form
\begin{eqnarray}
S&=&\frac{M_p^2}{2} \int dt d^3\bx N \sqrt{g}(K_{ij}
\mG^{ijkl}K_{kl}-\mV)\nonumber \\
&+&\frac{M_p^2}{2M^2_*}\int dt d^3\bx N\sqrt{g}
(D_i A_{jk}D_l A_{mn}M^{ijklmn}\nonumber \\
&+&2(\sigma_1 \mA_i \mA^i+\sigma_2 \mA_i D^i A+\sigma_3 \mA_i D_j A^{ij})+B^{ij}
(A_{ij}-K_{ij})) \ .
\nonumber \\
\end{eqnarray}
However we still see that there is the second problem with the
variable $\mA_i$ that contains the time derivative of
$a_i=\frac{\partial_i N}{N}$. In order to simplify given expression
let us begin with the following observations
\begin{equation}
\partial_t a_i=\partial_i\left(\frac{\partial_t N}{N}\right) \ , \quad
D_i a_j
=D_j a_i
\end{equation}
so that  $\mA_i$ can be written as
\begin{eqnarray}
\mA_i=
\frac{1}{2N}\partial_i \left(\frac{\partial_t
N}{N}-\frac{N^j}{N}\partial_j N\right) \ .
\nonumber \\
\end{eqnarray}
To proceed further we declare that $\mA_i$ is an independent
variable when we introduce auxiliary field $Y^i$ and add following
term to the action $Y^i\left(\mA_i-\frac{1}{2N}\partial_i
\left(\frac{\partial_t N}{N}-\frac{N^j}{N}\partial_j
N\right)\right)$. Then with the help of the integration by parts
\footnote{Since we are interested in the local properties of this
theory and the number of physical degrees of freedom we can ignore
boundary terms.}
  we can rewrite the action into the form
\begin{eqnarray}\label{Sfinal}
S&=&\frac{M_p^2}{2} \int dt d^3\bx N \sqrt{g}(K_{ij}
\mG^{ijkl}K_{kl}-\mV)\nonumber \\
&+&\frac{M_p^2}{2M^2_*}\int dt d^3\bx N\sqrt{g}
(D_i A_{jk}D_l A_{mn}M^{ijklmn}\nonumber \\
&+&2(\sigma_1 \mA_i \mA^i+\sigma_2 \mA_i D^i A+\sigma_3 \mA_i D_j A^{ij})+B^{ij}
(A_{ij}-K_{ij}))\nonumber \\
&+&\int dt d^3\bx  \left(N Y^i\mA_i+\frac{1}{2}\partial_i Y^i
\left(\frac{\partial_t N}{N}-\frac{N^j\partial_j N}{N}\right)\right)
\nonumber \\
\end{eqnarray}
which  is suitable for the Hamiltonian analysis. Explicitly  from
(\ref{Sfinal}) we obtain
\begin{eqnarray}
\pi^{ij}&=&\frac{\delta \mL}{\delta \partial_t g_{ij}}=
\frac{M_p^2}{2}\sqrt{g}\mG^{ijkl}K_{kl}
-\frac{M_p^2}{2M^2_*}\sqrt{g}B^{ij} \ , \quad \pi^i=\frac{\delta
L}{\delta
\partial_t N_i}
\approx 0 \ ,   \nonumber \\
p_{ij}&=&\frac{\delta S}{\delta \partial_t B^{ij}}\approx 0 \ , \quad
q^{ij}=\frac{\delta
S}{\delta \partial_t A_{ij}}\approx 0 \ , \nonumber \\
\mP^i&=&\frac{\delta \mL}{\delta \partial_t \mA_i}\approx 0 \ , \quad
p_i=\frac{\delta \mL}{\delta \partial_t Y^i}\approx 0 \ , \quad
\pi_N=\frac{1}{2N}\partial_iY^i \ ,   \nonumber \\
\end{eqnarray}
where the last relation implies following primary constraint
\begin{equation}
 \Phi_N=\pi_N N-\frac{1}{2}\partial_i Y^i
\approx 0 \ .
\end{equation}
Then the Hamiltonian has the form
\begin{eqnarray}
H&=&\int d^3\bx (\pi_N\partial_t N+\pi^{ij}\partial_t g_{ij}-\mL)=
\nonumber \\
&=&\int d^3\bx \left(N\mH_0+N^i\left(-2g_{ik}D_j\pi^{jk}
+\frac{1}{2N}\partial_i N\partial_j Y^j
\right)\right) \ , \nonumber \\
\end{eqnarray}
where
\begin{eqnarray}
\mH_0&=& \frac{2}{M_p^2\sqrt{g}}
(\pi^{ij}+\frac{M^2_p}{2M^2_*}\sqrt{g}B^{ij})\mG_{ijkl}(\pi^{kl}+
\frac{M^2_p}{2M^2_*}\sqrt{g}B^{kl})+\frac{M_p^2}{2}\sqrt{g}\mV-\nonumber \\
&-&\frac{M_p^2}{2M^2_*}\sqrt{g}
\left(D_i A_{jk}D_l A_{mn}M^{ijklmn} \right.\nonumber \\
&+& \left. 2(\sigma_1 \mA_i \mA^i+\sigma_2 \mA_i D^i A+\sigma_3
\mA_i D_j A^{ij})+B^{ij}
A_{ij}\right)- Y^i\mA_i \ .  \nonumber \\
\end{eqnarray}
Observe that with the help of the constraint $\Phi_N$ we can rewrite
an expression in the second bracket into the form
\begin{equation}
-2g_{ik}D_j\pi^{jk} +\frac{1}{2N}\partial_i N\partial_j Y^j=
-2g_{ik}D_j \pi^{jk}+\pi_N\partial_iN-N^i\partial_iN\Phi_N \ .
\end{equation}
Then it is easy to see that  an extended Hamiltonian has the form
\begin{eqnarray}
H_E=\int d^3\bx (N\mH_0+N^i\mH_i+v_N\Phi_N+\mP^iv^{\mP}_i+v^i_p
p_i+v^{ij}p_{ij}+w_{ij}q^{ij}) \ ,
\end{eqnarray}
where
\begin{equation}
\mH_i=-2g_{ik}D_l\pi^{kl}+\pi_N\partial_iN \ .
\end{equation}
As the next step we analyze the requirement of the preservation of
all primary constraints. We begin with $\pi_i\approx 0$
\begin{equation}
\partial_t\pi_i=\pb{\pi_i,H_E}=-\mH_i\approx 0 \ .
\end{equation}
However $\mH_i$ is not the correct form of the spatial
diffeomorphism constraints since it has vanishing Poisson brackets
with auxiliary fields. In order to find the correct form of the
spatial diffeomorphism constraints we add appropriate linear
combinations of the primary constraints to it so that we define
$\tmH_i$ as
\begin{eqnarray}
\tmH_i&=&\mH_i
 -2\partial_k (A_{ij} g^{jk}) +\partial_i A_{jk}q^{jk}
\nonumber \\
&+&2\partial_j (B^{jk}p_{ik})+
\partial_i B^{jk}p_{jk}- \partial_ip_jY^j+\partial_j(p_iY^j) \nonumber \\
\end{eqnarray}
and its smeared form
\begin{equation}
\bT_S(N^i)=\int d^3\bx N^i\tmH_i \ .
\end{equation}
Note that the extra terms that we added to $\mH_i$ are proportional
to the primary constraints. Then we obtain following Poisson
brackets
\begin{eqnarray}
\pb{\bT_S(N^i),A_{mn}}&=&-\partial_n N^i A_{im}-\partial_m N^i
A_{in} -N^i\partial_i A_{mn}
\ , \nonumber \\
\pb{\bT_S(N^i),B^{mn}}&=&\partial_j N^m B^{jn}+\partial_j N^n
B^{jm}-
N^i \partial_i B^{mn} \ , \nonumber \\
\pb{\bT_S(N^i),a_i}&=&-N^k\partial_k a_i-\partial_i N^k a_k \ , \nonumber \\
\pb{\bT_S(N^i),Y^i}&=&-\partial_k(N^k Y^i)+\partial_j N^i Y^j \ .
\nonumber
\\
\end{eqnarray}
Note that $Y^i$ transforms as a vector density.  These relations
show that $\tmH_i$ are correct generators of spatial diffeomorphism.
Further, they are preserved during the time evolution of the system
due to the fact that the Hamiltonian is manifestly invariant under
spatial diffeomorphism and also due to the fact that the Poisson
brackets between the smeared form of these constraints is equal to
\begin{equation}
\pb{\bT_S(N^i),\bT_S(M^j)}= \bT_S(N^j \partial_j M^i-M^j\partial_j
N^i) \ .
\end{equation}
Now we consider time evolution of the constraint $\mP^i\approx 0$
\begin{eqnarray}
\partial_t \mP^i&=&\pb{\mP^i,H}=N\left(\frac{M_p^2}{M^2_*}\sqrt{g}(2\sigma_1
\mA^i+\sigma_2 D^iA+\sigma_3 D_j A^{ij})+Y^i\right)\equiv
N\Phi^i_{II}\approx 0 \
\nonumber \\
\end{eqnarray}
that implies an existence of the secondary constraints
$\Phi^i_{II}\approx 0$.
Let us now analyze the time evolution of the constraint
$\Phi_N\approx 0$
\begin{eqnarray}\label{timePhi}
\partial_t \Phi_N=\pb{\Phi_N,H}=
\pb{\Phi_N(\bx),\int d^3\by N(\by)\mH_0(\by)}
+\frac{1}{2}\partial_i v^i=0 \ .  \nonumber \\
\end{eqnarray}
Naively we should say that this equation determines the Lagrange
multipliers $v^i$ which is not correct since it would imply that one
equation determines three components of $v^i$. In order to resolve
this issue note that we can replace $v^i$ with arbitrary
combinations of another Lagrange multipliers and hence let us
presume  $v^i$ in the form  $v^i=\epsilon^{ijk}\partial_j \tv_k$
that obeys $\partial_i v^i=0$. With this ansatz the equation
(\ref{timePhi}) implies an existence of the secondary constraint
$\mC$.  In order to find its explicit form we use the fact that
\begin{equation}
\pb{\pi_N(\bx),a_i(\by)}=\frac{1}{N}\partial_i \delta(\bx-\by) \
\end{equation}
and hence
\begin{equation}
\pb{\Phi_N(\bx),\int d^3\by N(\by)\mH_0(\by)}
-N(\bx)\mH_0(\bx)+\frac{\partial}{\partial x^i} \left(\frac{\delta
\mH_0}{\delta a_i(\bx)}\right) \equiv -N\mC(\bx) \ ,
\end{equation}
where $\mC=\mH_0-\frac{1}{N}\partial_i\left[\frac{\delta \mH_0}
{\delta a_i}\right]$. Note that this constraint is the
generalization of the constraint $\mC$ known from the
non-projectable HL gravity
\cite{Chaichian:2015asa,Kluson:2010nf,Donnelly:2011df} to its mixed
derivative generalization.

It is also important to stress that when we replace the Lagrange
multipliers $v^i$ with $\tv^i$ we find that after integration by
parts the original  constraints $p_i\approx 0$ are replaced with
equivalent ones
\begin{equation}
\psi^i_p=\epsilon^{ijk}\partial_j p_k\approx 0 \ .
\end{equation}
Now the requirement of the preservation of the constraint $\psi^i_p\approx 0$
implies
\begin{equation}
\partial_t \psi^i_p=\pb{\psi^i_p,H}=\epsilon^{ijk}\partial_j (N\mA_k)-\frac{1}{2}\epsilon^{ijk}\partial_j\partial_k v_N=
\epsilon^{ijk}\partial_j (N\mA_k)\equiv \Psi^i_{II}\approx 0 \ .
\end{equation}
Note that $\psi^i_p$ and $\Psi^i_{II}$ are not independent but obey the relation
\begin{equation}
\partial_i\psi^i_p=0 \ , \partial_i\Psi^i_{II}=0 \ .
\end{equation}
 This is very important result which will be useful when we calculate the number
of physical degrees of freedom.

Now we analyze the requirement of the preservation of the
constraints $p_{ij}\approx 0$
\begin{eqnarray}
\partial_t p_{ij}=\pb{p_{ij},H_E}=-N\frac{4}{M_p^2} \mG_{ijkl}
(\pi^{kl}+\frac{1}{2}\sqrt{g}B^{kl})-N\sqrt{g}A_{ij}\equiv -N\Phi_{ij}^{II}\approx 0 \ , \nonumber \\
\end{eqnarray}
and $q^{ij}\approx 0$
\begin{eqnarray}
\partial_t q^{ij}&=&\pb{Q^{ij},H_E}
=N\left(\frac{M_p^2}{M^2_*}\sqrt{g} a_k M^{kijlmn}D_l
A_{mn}-\frac{M_p^2}{M^2_*}\sqrt{g}\sigma_2
a^k \mA_k g^{ij}\right.\nonumber \\
&-&\frac{M_p^2}{2M_*^2}(a^i \mA^j+a^j\mA^i)-\sqrt{g}B^{ij}
+\frac{M_p^2}{M^2_*}\sqrt{g} D_k (M^{kijlmn}D_l
A_{mn})-\frac{M_p^2}{M^2_*}\sqrt{g}\sigma_2
D^k (\mA_k) g^{ij}\nonumber \\
&
&\left.-\frac{M_p^2}{2M_*^2}(D^i(\mA^j)+D^j(\mA^i))-\sqrt{g}B^{ij}\right)\equiv
N\Psi^{ij}_{II}\approx 0  \ . \nonumber \\
\end{eqnarray}
It is easy to see that $\Phi^{II}_{ij},\Psi_{II}^{ij}$ are the
second class constraints with $q^{ij}\approx 0 \ , p_{ij}\approx 0$
so that they vanish strongly and should be solved at least in
principle.
 Further, these
constraints do not depend on $N$ which is also very important.

In summary we have following collection of the second class
constraints
\begin{equation}
\psi^i_p\approx 0 \ , \quad  \Psi^i_{II}\approx 0 \ , \quad
\mP^i\approx 0 \ , \quad  \Phi^i_{II} \approx 0 \ , \quad
\Phi_N\approx 0 \ , \quad  \mC\approx 0 \ .
\end{equation}
From $\Phi^i_{II}$ we can express $\mA_i$ as function of the
canonical variables at least in principle. Using $\mC$ and $\Phi_N$
we can eliminate $\pi_N$ and $N$. Finally note that $\psi^i_p$ and
$\Psi^i_{II}$ can be solved  as
\begin{equation}
p_i=\partial_i \pi \ , N\mA_i=\partial_i\phi
\end{equation}
which implies an existence of one scalar degree of freedom $\phi$
with momenta $\pi$. This result also explains why we used the
Lagrange multipliers $\tv_i$ instead of $v^i$. The goal was to
separate the second class constraints into two ones $\mC,\Phi_N$
that eliminate $N$ and conjugate momenta and the remaining ones that
implies an existence of the scalar degree of freedom. Finally note
that $\Phi^{II}_{ij}$ and $\Psi^{ij}_{II}$ can be solved for
$A_{ij}$ and $B^{ij}$ at least in principle so that there are no new
additional dynamical degrees of freedom. This ie expected result
since the spatial derivative of $K_{ij}$ cannot generate new
dynamical degrees of freedom. Finally note that three first class
constraints $\tmH_i$ can be gauge fixed and we eliminate three
degrees of freedom from $g_{ij}$. The remaining three degrees of
freedom correspond to the massless graviton and one scalar degree of
freedom. In summary, mixed derivative HL gravity contains two
additional  scalar degrees of freedom with respect to GR.

As the last point we proceed to the question of an existence of the
global constraints. Note that the action is invariant under
foliation preserving diffeomorphism and hence we expect an existence
of two global first class constraints as in case of non-projectable
HL gravity \cite{Donnelly:2011df}. Let us introduce the first one
\begin{equation}
\Pi_N=\int d^3\bx \Phi_N=\int d^3\bx \pi_N N
\end{equation}
that has following  Poisson brackets
\begin{equation}
\pb{\Pi_N,N}=-N \ , \quad  \pb{\Pi_N,\pi_N}=\pi_N \ , \quad
\pb{\Pi_N,a_i}=0 \
\end{equation}
together with
\begin{equation}
\pb{\Pi_N,\Psi^i_{II}}=-\epsilon^{ijk}\partial_j (N\mA_k)=-\Psi^i_{II}
\approx 0 \ .
\end{equation}
In other words $\Pi_N$ poisson commutes with all second class
constraints that do not  depend on $N$ explicitly. The situation is
slightly more complicated in case of the constraint $\mC(\bx)$ we
use following result
\begin{eqnarray}
& &\pb{\Pi_N,\mC(\bx)}=\pb{\Pi_N,\pb{\Phi_N(\bx),H_E}}=\nonumber \\
&=&\pb{\Phi_N(\bx),\pb{\Pi_N,H_E}}=
-\pb{\Phi_N(\bx),\int d^3\by N(\by) \mH_0(\by)}=N\mC(\bx)\approx 0 \
.  \nonumber
\\
\end{eqnarray}
In summary we found that $\Pi_N$ has vanishing Poisson bracket with
all second class constraints on the constraint surface.

 Then, following
\cite{Chaichian:2015asa}
 we  split $\Phi_N$ as
\begin{equation}
\tPhi_N=\Phi_N-\frac{\sqrt{g}N}{\int d^3\bx \sqrt{g}N}\Pi_N \ ,
\end{equation}
that obeys the equation
\begin{equation}
\int d^3\bx \tPhi_N=0 \ .
\end{equation}
In other words we have $\infty^3-1$ local second class constraints
$\tPhi_N$ and one global first class constraint $\Pi_N$ \footnote{We
use notations introduced in \cite{Kuchar:1991xd}. It is important to
stress that $\tPhi_N$ and $\Pi_N$ consist $\infty^3$ constraints
which corresponds to the number of the constraints  $\Phi_N$.}.
Finally the requirement of the preservation of the constraint
$\Pi_N$ implies
\begin{equation}
\partial_t\Pi_N=\pb{\Pi_N,H_E}=-\int d^3\bx N\mH_0\equiv -\Pi_T\approx
0 \ ,
\end{equation}
where  we introduced new first class constraint $\Pi_T\approx 0$ and
we used the fact that $\mH_0$ depends on $a_i$ only and hence
Poisson commutes with $\Pi_N$. Observe that using the explicit form
of $\mC$ we  have the relation
\begin{equation}
\int d^3\bx N\mC=\int d^3\bx N\mH_0 \ .
\end{equation}
  Then we can introduce the
constraint
\begin{equation}
\tmC(\bx)=\mC(\bx)-\frac{\sqrt{g}}{\int d^3\bx
\sqrt{g}N}\Pi_T
\end{equation}
that obeys the relation
\begin{equation}
\int d^3\bx N\tmC=0 \
\end{equation}
so that there is  $\infty^3-1$  second class constraints $\tmC$.

 As
we argued previously $\Pi_N$ is the first class constraint. Further,
since $\mH_0$ does not depend on $N$ but on $a_i$ only we find that
\begin{equation}
\pb{\Pi_N,\Pi_T}=0 \ .
\end{equation}
On the other hand it is clear that $\Pi_T$ does not have vanishing
Poisson brackets with all second class constraints and hence we
cannot say that it is the first class constraint. In order to
resolve this problem we proceed as follows. As the first step
 we introduce common notation for all  second class constraints
  $\Psi_A$  and denote
 their Poisson brackets as
 \begin{equation}
\pb{\Psi_A(\bx),\Psi_B(\by)}=\triangle_{AB}(\bx,\by)
\end{equation}
with inverse
\begin{equation}
\int d^3\by \triangle_{AB}(\bx,\by)\triangle^{BC}(\by,\bz)=
\delta_A^C\delta(\bx,\bz) \ .
\end{equation}
Then we define following constraint
\begin{equation}
\tPi_T=\Pi_T-\int d^3\bx
d^3\by\pb{\Pi_T,\Psi_A(\bx)}\triangle^{AB}(\bx,\by)\Psi_B(\by)
\end{equation}
that clearly obeys the relation
\begin{equation}
\pb{\tPi_T,\Pi_N}=0 \ , \pb{\tPi_T,\Psi_A(\bx)}=0
\end{equation}
and also
\begin{eqnarray}
\pb{\tPi_T,\tPi_T}=0 \ . \nonumber \\
\end{eqnarray}
In summary we have found the second first class constraint $\tPi_T$
that reflects the invariance of the action under foliation
preserving diffeomorphism. Of course, this constraint reduces to
$\Pi_T$ when all second class constraints vanish strongly.
\section{Summary}\label{fourth}
This short note was devoted to the Hamiltonian analysis of the mixed
derivative extension of HL gravity. We showed that there is a new
scalar mode with agreement with the perturbative analysis performed
in \cite{Coates:2016zvg}. In other words mixed derivative HL gravity
contains two additional scalar modes with respect to the GR. The
presence of these modes could have huge impact on the consistency of
the theory and on its phenomenological applications as was discussed
in \cite{Coates:2016zvg}.

Naively we should say that when we add terms with time derivative of
the lapse $N$ then the presence of the new dynamical degree of
freedom is obvious but the situation is not so simple. The reason is
that $N$ is not ordinary scalar but it transforms under foliation
preserving diffeomorphism as
\begin{equation}
N'(t',\bx')=(1-f(t))N(t,\bx) \ .
\end{equation}
Than the naive time derivative $\nabla_n N=\frac{1}{N}(\partial_t
N-N^i\partial_i N)$ is not invariant under foliation preserving
diffeomorphism and it turns out that the only possible  covariant
expression that contains time derivative of $N$ is  a vector $\mA_i$
and hence terms that are presented in $\mL_\kappa$ that contain
mixed derivatives. As we saw in the main body of this paper these
terms  make the Hamiltonian analysis rather non-trivial.

The second important point was to identify two global first class
constraints  whose existence is predicted by the fact that the mixed
derivative HL gravity is invariant under foliation preserving
diffeomorphism.  We found these constraints and we showed that they
are the first class constraints.

We can also comment the issue of two additional degrees of freedom.
In principle they could be eliminated when we introduce some
Lagrange multiplier modified terms to the action as for example in
\cite{Chaichian:2015asa}. These new second class constraints could
eliminate these additional degrees of freedom at least in principle.
However it is important to stress that the resulting theory  will be
very complicated. Even without these additional terms the mixed
derivative HL gravity has very complicated symplectic structure due
to the fact that there are second class constraints that contain
differential operators. These facts together with the existence of
two scalar degrees of freedom is very important problem  when we
consider mixed derivative HL gravity as a candidate of  the
renormalizable theory of gravity.

 \noindent {\bf Acknowledgements}
\\
 This  work  was supported by the Grant Agency of the
Czech Republic under the grant P201/12/G028.


\end{document}